\documentclass[11pt]{article}

\usepackage{amsfonts,amsmath,amsthm}

\usepackage{hhline}
\usepackage{hyperref}
\usepackage[english]{babel}
\usepackage[utf8]{inputenc}
\usepackage{url}
\usepackage{graphicx}
\usepackage{textcomp}
\usepackage{floatflt}
\usepackage[font={footnotesize}]{caption}
\usepackage{subcaption}
\usepackage{color}
\usepackage{makecell}
\usepackage{algpseudocode}
\usepackage{algorithm}
\usepackage{geometry}
\definecolor{szin}{rgb}{0,0.44,0.4}
\definecolor{szin2}{rgb}{0.902,0.2705,0}
\definecolor{szin3}{rgb}{0,0.5,0}

\hypersetup{pdfpagemode=UseNone, pdftoolbar=true, breaklinks=true,
colorlinks=true,
linkcolor=szin2,
linktoc=page,
citecolor=szin3,bookmarks=false}

\theoremstyle{defin}
\newtheorem*{defin}{Definition}

\begin{document}

\title{\huge Co-occurrence simplicial complexes in mathematics: identifying the holes of knowledge}


\author{{
\sc Vsevolod Salnikov}$^*$,\\[2pt]
NaXys, Universit\'{e} de Namur, 5000 Namur, Belgium\\
$^*${\texttt{Corresponding author: vsevolod.salnikov@unamur.be}}\\[2pt]
{\sc Daniele Cassese} \\[2pt]
NaXys, Universit\'{e} de Namur, 5000 Namur, Belgium\\
ICTEAM, Universit\'{e} Catholique de Louvain, 1348 Louvain-la-Neuve, Belgium\\
Mathematical Institute, University of Oxford, OX2 6GG Oxford, UK\\
{daniele.cassese@unamur.be}\\[6pt]
{\sc Renaud Lambiotte}\\[2pt]
Mathematical Institute, University of Oxford, OX2 6GG Oxford, UK\\
{renaud.lambiotte@maths.ox.ac.uk}\\
{\sc and}\\[6pt]
{\sc Nick S. Jones}\\[2pt]
Department of Mathematics, Imperial College, SW7 2AZ London, UK\\
{nick.jones@imperial.ac.uk}}
\date{}

\maketitle

\begin{abstract}
{In the last years complex networks tools contributed to provide insights on the structure of research, through the study of collaboration, citation and co-occurrence networks. The network approach focuses on pairwise relationships, often compressing multidimensional data structures and inevitably losing information. In this paper we propose for the first time a simplicial complex approach to word co-occurrences, providing a natural framework for the study of higher-order relations in the space of scientific knowledge. Using topological  methods we explore the conceptual landscape of mathematical research, focusing on homological holes, regions with low connectivity in the simplicial structure. We find that homological holes are ubiquitous, which suggests that they capture some essential feature of research practice in mathematics. Holes die when a subset of their concepts appear in the same article, hence their death may be a sign of the creation of new knowledge, as we show with some examples. We find a positive relation between the dimension of a hole and the time it takes to be closed: larger holes may represent potential for important advances in the field because they separate conceptually distant areas. We also show that authors' conceptual entropy is positively related with their contribution to homological holes, suggesting that \emph{polymaths} tend to be on the frontier of research. }\\[2mm]

\noindent \textbf{Keywords:} Co-occurrence, Topological Data Analysis, Persistent Homology, Simplicial Complexes
\\[1mm]
\end{abstract}

\section{Introduction}

Co-occurrence networks capture relationships between words appearing in the same unit of text: each node is a word, or a group of words, and an edge is defined between two nodes if they appear in the same unit of text. Co-occurrence networks have been used, among other things, to study the structure of human languages \cite{Ferrer01}, to detect influential text segments \cite{Garg18} and to identify authorship signature in temporal evolving networks \cite{Akimushkin17}. Other applications include the study of co-citations of patents \cite{Wang11}, articles \cite{Lazer09} and genes \cite{Jenssen01, Mullen14}.  Here we focus on the co-occurrences of concepts (theorems, lemmas,  equations) in scientific articles to gain understanding in  the structure of knowledge in Mathematics. Similar problems have been considered in scientometrics, even if  previous works have  limited their analysis to  keywords, or words appearing in  abstracts  \cite{Radhakrishnan17, Zhang12, Su10}, and focused only on binary relations between words, as we clarify below.  

The main novelty of our work is to study co-occurrences in a simplicial complex framework, using persistent homology to understand the conceptual landscape of mathematics.
The adoption of a simplicial complex framework is motivated by the fact that concepts are inherently hierarchical, so simplicial complexes might seem a natural representation: often elementary conceptual units connect together to form nested sequences of higher-order concepts. A simplicial complex approach to model the semantic space of concepts was already suggested by \cite{Chiang07}, even if not in a topological data analysis framework\cite{Patania17_2}, while application of topological data analysis tools to visualisation of natural language can be find in  \cite{Jo11, Wagner12, Sami17}. 
Several reasons motivate the use of higher-order methods in this context. First, co-occurrence networks tend to be extremely dense in practice and require additional tools to filter the relations and sparsify the network to extract information \cite{Serrano09, Slater09}. Second, in the original dataset, interactions are not pairwise and it is unclear if the constraints induced by a network framework, in terms of nodes and pairwise edges, do not obscure important structures in the system.
By modelling co-occurrence relations as a simplicial complex, we thus go beyond the network description that reduces all the structural properties to pairwise interactions and their combinations, explicitly introducing higher-order relations. Note that this modelling approach, in particular the use of simplicial persistent homology, has found uses when the data is inherently multidimensional \cite{Petri13}, with applications in neuroscience \cite{Petri14, Stolz17}, biology \cite{Chan13, Mamuye16} to the study of contagion \cite{Taylor15} and to coauthorship networks \cite{Patania17, Carstens13}.

A second contribution of our work is the analysis of the full text of a large corpus of articles, which allows us  to bypass the high-level categorisation provided by keywords but also to identify the use of methodological tools and to gain insight into mathematical praxis. However, the main purpose of this article is to use the resulting dataset of concepts and articles as a testbed in which to apply methods from topological data analysis, and to go beyond a standard network analysis. 

\section*{Dataset}
The dataset analysed has been scraped from arXiv, and includes a total of 54177 articles  from 01/1994 to 03/2007, of which 48240 in mathematics (\texttt{math}) and 5937 in mathematical physics (\texttt{math-ph}). We have limited the timeframe due to naming conventions in arXiv: since 03/2007 subject is not a part of the article identifier, thus if one wants to export it additional queries to metadata are needed. That is easily expandable, but we decided to limit the dataset at this moment for computational speed. The date is extracted from article id, hence it refers to the submission date. Notice that some of articles in the first years may have been written some years before 1991 (arXiv first article's date). In order to describe the mathematical content of articles from the \LaTeX \ file we look at different concepts occurrences in the text. Clearly the choice of the concepts set can influence the outcome: choosing them manually by a small group of people would result in a strong bias towards the understanding and priorities of the individuals in the group. Thus we wanted to have something either globally accepted by scientific community or at least created by a sufficiently large group of people. Another point in the selection of a good concepts list is the possibility to make a similar research for other disciplines, thus we chose to get it from some general, easily accessible source. Our strategy consisted in parsing a concepts list from Wikipedia, which includes 1612  equations, theorems, lemmas. Clearly these concepts are not homogeneous, meaning that some of them might represent extremely specific theorems, while others can be very general, like \emph{differential equation}, but similar holds for any text processing with different words having different frequencies. Our position on that is still to minimize the manually introduced bias: we consider that all concepts have similar weight and try to have as complete set as possible. Moreover it is possible that two different names represent for example the same theorem due to historical reasons. For the moment we consider such synonyms as distinct entities as the usage of one of them but not the other may reflect  structural properties: for example a  lemma might have  different names depending on a (sub-)field of mathematics and manually merging them is not correct.

\begin{table}[h!]
\centering
\caption{Dataset}
      \begin{tabular}{ccccc}
        \hline
         & Years &  Papers & Concepts   & Authors\\ \hline
      Total  & 1994 - 2007 & 35018 &  1612 & 23471 \\
       Included & 1994 - 2007 & 8375 & 1067 & 8852  \\
                 \hline
      \end{tabular}
\end{table}

\begin{figure}[h!]
  \includegraphics[width=.45\textwidth]{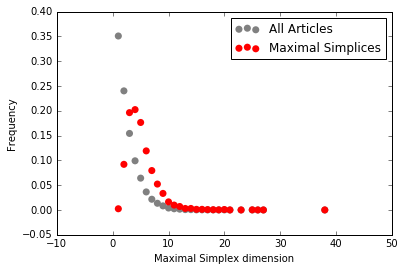}
  \includegraphics[width=.45\textwidth]{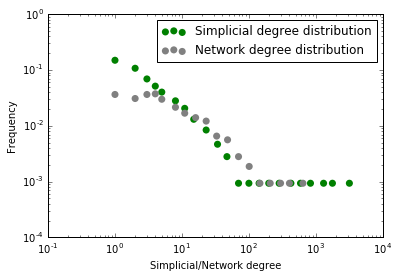}
  \caption{\textbf{Dimension of maximal simplices and simplicial degree}
     On the lefthand side the distribution of number of concepts per article. Here we compare the distribution of the entire dataset (grey) with that of the papers included in the simplicial complex, which shows that most of the articles with few concepts in are not included as they are a subset of other articles. On the righthand side the simplicial and network degree distribution are showed.}
      \label{MSC}
      \end{figure}

As a next step, we combine both datasets.
Among the whole concepts list, 1067 find a match in at least one article. Among the 54177 articles, 35018 contain at least one of the concepts in our list (30369 for mathematics and 4649 for mathematical physics), and we also take the list of authors to analyse their contribution to the conceptual space. We construct the binary (non-weighted) co-occurrence simplicial complex (defined more formally below) over the 1067 nodes by including a $(k-1)$-simplex for each article containing $k$ concepts, provided its concept set is not fully included in the concept set of another article, that is we only keep \emph{facets} of the simplicial complex. Whenever the concept sets of two articles intersect, their corresponding simplices share a face of dimension $(n-1)$, where $n$ is the dimension of the intersection.
The corresponding network, namely the  1-skeleton of the co-occurrence simplicial complexes (that is we only look at faces of dimension 0 and 1) has 1067 nodes and 32707 unweighted edges. Figure \ref{MSC} shows the network and simplicial (concept) degree distribution, where the simplicial degree of a concept is the number of \emph{facets} (articles) it belongs to. The sum of all simplicial degrees is 42009, which means there are 39.37 concepts per paper on average.

\section*{Simplicial Complexes}
A simplicial complex is a space obtained as the union of simple elements (nodes, edges, triangles, tetrahedra and higher dimensional polytopes). Its elements are called simplices, where a $k$-simplex is a set of $k+1$ elements with the property that each of its subsets is also a simplex. Subsets of a simplex of dimension $d$ are called \emph{d-faces}, and two simplices intersect if they share a common face.
More formally:
\begin{defin}
A $n$-dimensional simplex in $\mathbb{R}^m$ is the convex hull of $n+1$ points in general position in $\mathbb{R}^m$, with $m \ge n$,
\end{defin}
We emphasize here, that the metrics inside $\mathbb{R}^m$ is not used here and this definition can be replaced to a combinatorial one without any further changes.

\begin{defin}
A Simplicial Complex in $\mathbb{R}^m$ is a family $X$ of simplices in $\mathbb{R}^m$ such that:
\begin{itemize}
\item $\sigma \in X$ and $\tau \subset \sigma$ then $\tau \in X$
\item $\sigma^1 \cap \sigma^2 \ne  \emptyset$ or $\sigma^1 \cap \sigma^2 \subseteq \sigma^1$ and $\sigma^1 \cap \sigma^2 \subseteq \sigma^2$
\end{itemize}
\end{defin}

Simplicial complexes can be seen as generalisation of a network beyond pairwise interactions, that differ from hypergraphs as all subsets of a simplicial complex must also be simplices.
As an illustrative example of how simplicial complexes capture higher-order interactions where networks fail to do so, consider that in a co-occurrence network it is not possible to distinguish between three concepts appearing in the same paper and three concepts appearing in three papers each containing two concepts: in a network both cases are represented  by a triangle, while in a simplicial complex the first is a 2-simplex (a filled triangle) and the second is a cycle made of three 1-simplices (an empty triangle).

 \begin{figure}[h!]
 \centering
 \includegraphics[width=5cm]{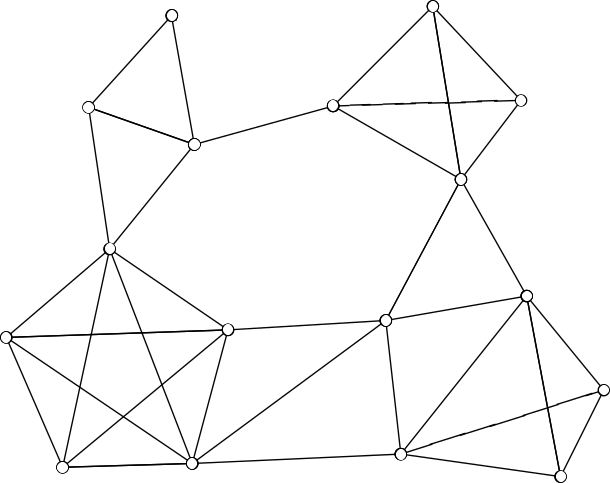}
  \includegraphics[width=5cm]{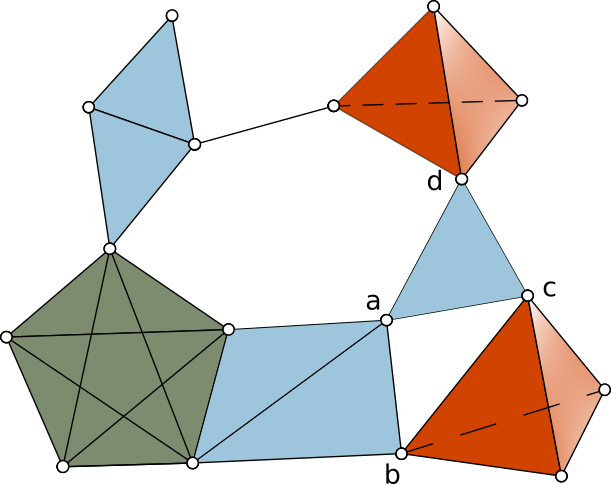}
  \caption{\textbf{Network (left) and simplicial complex (right) representation.}
       The (maximal) 2-simplices are in light blue, 3-simplices in red and the 4-simplex is in green. Notice that in the network these are indistinguishable.}
      \label{SC}
      \end{figure}

\begin{figure}[h!]
        \centering
 \includegraphics[width=.5\textwidth]{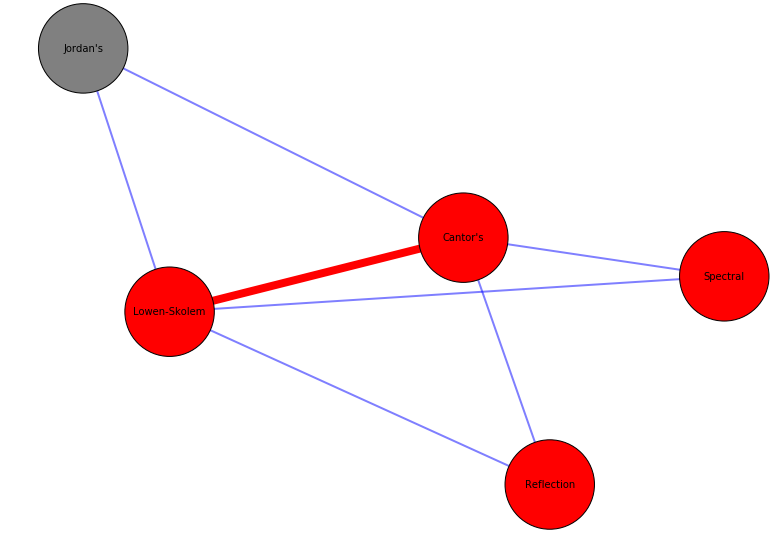}\\
  \includegraphics[width=.5\textwidth]{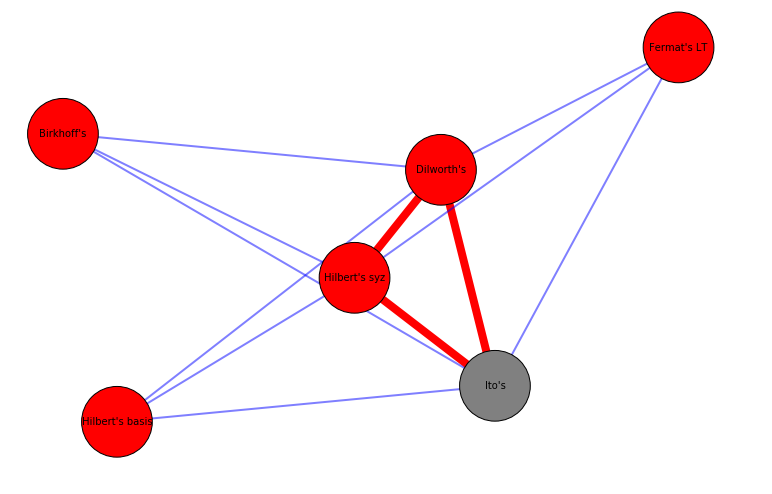}
  \caption{\textbf{Simplicial stars}
      Example of a $S^3_3$ (top) and a $S^4_3$ (bottom) star. Red nodes are theorem, grey nodes lemmas, while red edges delimit the simplices in core of the star.}
      \label{stars}
      \end{figure}

As for networks, also for simplicial complexes we can define simplicial measures that are the higher-order analogs of networks ones, for example  \cite{Estrada17} defines several simplicial centrality measures, providing also the characterisation of some families of simplicial complexes. In this paper we use the simplicial analogs of stars to provide a further description of the concept space.
\begin{defin}
A simplicial star $S_l^k$ consists of a central $(k-1)$-simplex that is a face of $l$ $k$-simplices, and there is no other simplex but their subsimplices. 
\end{defin}
$S^1_5$ for example is the usual star, with a node in the core connected to 5 nodes, while for $S^3_5$ the core is an edge. Examples are reported in Figure \ref{stars}.

\section*{Persistent Homology}

Persistent homology is a method in topological data analysis  \cite{Carlsson09, Patania17_2} based on algebraic topology, that studies the shape of the data by finding holes of different dimensions in the dataset. Holes are topological invariants that can be seen as voids bounded by simplices of different dimension: in dimension 0 they are connected components, in dimension 1 loops (voids bounded by edges), in dimension 2 voids bounded by triangles and so on. 
Here we give a brief and intuitive explanation of the homology of simplicial complexes, for details on how to compute it we refer to \cite{Edelsbrunner08, Horak09, Otter17}.

For every $k$-simplex of a simplicial complex $K$, consider the simplicial analog of a path, a $k$-chain, simply the formal sum of adjacent $k$-simplices (where by adjacent we mean that they share one $k-1$ face) with coefficients  in some algebraic ring $R$ uniquely identifying the chain (for example a 2-chain is a formal sum of oriented edges). It is a common practice to consider $Z$ or $Z_n$ for $R$ and negative coefficients change the orientation, inherited from $\mathbb{R}^m$. Without any limitations and for the sake of simplicity one can imagine $R=Z_2$ as it permits us to eliminate questions of orientation: in this case $-1 S = S$ for any simplex $S$. If we consider a  $k$-simplex, it is bounded by its $(k-1)$ faces, and we call the corresponding $(k-1)$-chain, equal to the sum of these faces with coefficients 1 and coherent orientations, the boundary of that simplex. Again to make an illustrative example, the boundary of a 2-simplex (filled triangle) is then a formal sum of its 1-faces (edges). 
The boundary for a general $k$-chain is defined as the sum of the boundaries of the simplices in the chain taken with corresponding coefficients. Consider the linear map on the space of $k$-simplices, mapping each $k$-simplex to its boundary, the \emph{boundary operator}

\[
\partial_k : C_k (K) \rightarrow C_{k-1} (K)
\]
defined on the vector space with basis given by the simplices of $K$.
A $k$-cycle is defined as a $k$-chain without a boundary, hence it is an element of the kernel of $\partial_k$, and a $k$-boundary is a $k$-chain which is the boundary of a $k+1$ chain, so it is an element of the image of $\partial_{k+1}$, which is a subset of the kernel of $\partial_k$ as the boundary of a boundary is empty, or $\partial_{k}\partial_{k+1} = 0$.

So we have defined two interesting subspaces: the collection of $k$-cycles and the collection of $k$-boundaries, and we can also take the quotient space as the second is a subset of the first: what is left in the quotient space are those $k$-cycles that do not bound $(k+1)$-subcomplexes, and these are the $k$-dimensional voids. More precisely, as there can be more $k$-cycles around the same hole, the elements of the quotient space can be divided in homological classes, each identifying a hole. This quotient space is the $k$th homology of the simplicial complex

\[
H_k (K) = \frac{\text{ker}(\partial_k)}{\text{Im}(\partial_{k+1})}
\]

and its dimension 

\[
\beta_k (K)  = \text{dim ker}(\partial_k) - \text{dim Im}(\partial_{k+1})
\]

is the number of \emph{homology classes} or $k$-dimensional voids in the simplicial complex, the $k$-th \emph{Betti number} of the homology. 
For example the zero\emph{th} Betti number counts the number of connected components in the graph that constitutes the 1-skeleton of the simplicial complex, the first Betti number the number of loops, the second Betti number counts voids.

The homological features of  complexes are usually studied on a filtration of the complex, that is a sequence of simplicial complexes starting at the empty complex and ending with the full complex, so that the complex at step $n<m$ is embedded in the complex at $m$ for all the steps. In this way it is possible to focus on the persistency of homological features: as the filtration evolves the shape of data changes, so \emph{birth} and \emph{death} of holes can be recorded. A hole is born at step $s$ if it appears for the first time in the corresponding step of the filtration, and dies at $t$ if after step $t$ the hole disappears. The difference between birth and death of homological features is called \emph{persistence}, and can be recorded by a \emph{barcode}, a multiset of intervals bounded below \cite{Carlsson05} visualizing the lifetime of the feature and its location across the filtration: the endpoints of each interval are the step of the filtration where the homological feature is born and dies \cite{Horak09}.
An alternative visualisation is provided by the \emph{persistence diagrams}, which are built by constructing a peak function for each barcode, proportional to its length \cite{Stolz17}.

The way the filtration is built depends on the analysis that one wants to do on the data, a very common method on a weighted network is the \emph{weighted rank clique filtration} \cite{Petri13}. This is done by filtering for weights: after listing all edge weights $w_t$ in descending order, at every step $t$ one takes the graph obtained keeping  all the edges which weight is greater than or equal to $w_t$. The simplicial complex at that step of the filtration is built by including all the maximal $k$-cliques of the graphs as $k$-simplices. The obtained simplicial complex is called \emph{clique complex}.

In this paper we use a \emph{temporal filtration} instead, as in \cite{Pal17}. Using article dates we build a temporal filtration 

\[
\mathcal{F}_0 \subseteq \mathcal{F}_1 \subseteq \dots \subseteq \mathcal{F}_T
\] 

where $(0,T)$ are first and last date in our dataset (time step is one month)  and  for $i<j$  and each $\mathcal{F}_i$  co-occurrence complex contains  simplices of concepts (articles) up to date $i$. As every article is a simplex we do not need to build a clique complex like in the weighted rank clique filtration.

\section*{Reducing the computational burden}

Computing persistent homology is very costly if there are large simplices in the simplicial complex, as for each simplex the computation requires to list all the possible subsimplices. For instance in our (rather small) dataset, there are already simplices with 37 vertices and the number of its (k-1)-subsimplices is $\frac{37!} {(37-k!)k!}$, making it impossible to finish the task in reasonable time with standard tools. In order to reduce the computational burden, we put an upper bound on the dimension of simplices, that is we take the subcomplex that only includes simplices up to a maximum dimension $d_M =5$. In other words, we compute the homology of the $d_M$-\emph{skeleton} of our simplicial complex $K$.

In the $d_M$-skeleton of $K$ all simplices of dimension $d > d_M$ are replaced by collections of their ${d_M}$-faces, that is a complex of dimension $d_M$ made by glueing together  $\binom{d+1}{d_M+1}$  ${d_M}$-simplices along their  ${d_M}-1$ faces, such that for each  ${d_M}-1$ face there are $d -  {d_M}$ simplices sharing that face.
To make an illustrative example if ${d_M}=2$, the 2-skeleton of a 3-simplex is the collection of triangles in the boundary of the tetrahedron.

\begin{figure}[h!]
  \centering
 \includegraphics[width=10cm]{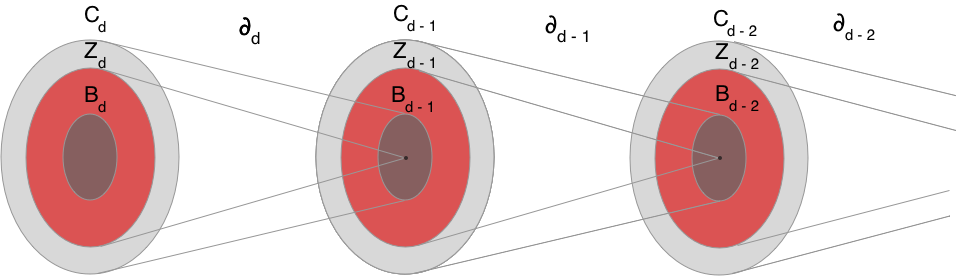}
  \caption{\textbf{Chains, cycles and boundaries sets}
       and their maps under the boundary operator for the $d$-skeleton of a complex $K$, with $d = d_M$. $C_d$, $Z_d$ and $B_d$ represent the collections of $d$ chains, cycles and boundaries respectively. Notice that in $K$ we may well have $C_{d+1}$ and bigger dimensional chains with corresponding boundary operators, while in $K^{d_M}$ we start from $C_{d_M}$ as the largest simplices we have are $d_M$-dimensional.}
      \label{maps}
      \end{figure}

It is straightforward to show that the $d_M$-skeleton of $K$, $K^{d_M}$ is $(d_M - 1)$-homologically equivalent to $K$, in the sense that they have the same homology groups up to $H_{(d_M - 1)}$.  Moreover, for $d \le d_M $, the $d$-chains group of the $d_M$-skeleton coincides with the $d$-chains group of $K$, as they have the same $d_M$-simplices, hence it follows that also $d$-cycles groups coincide (see Figure \ref{maps}). This implies that any map $\partial_d$ with $d \le d_M$ is the same on $K^{d_M}$ and $K$, hence the set of $d$-boundaries with $d \le d_M -1$ is the same on the two complexes. So  $H_d (K) =  \frac{\text{ker}(\partial_d)}{\text{Im}(\partial_{d+1})} = H_d (K^{d_M}) $ for $d \le d_M -1$.

\begin{figure}[h!]
        \centering
  \includegraphics[width=5cm]{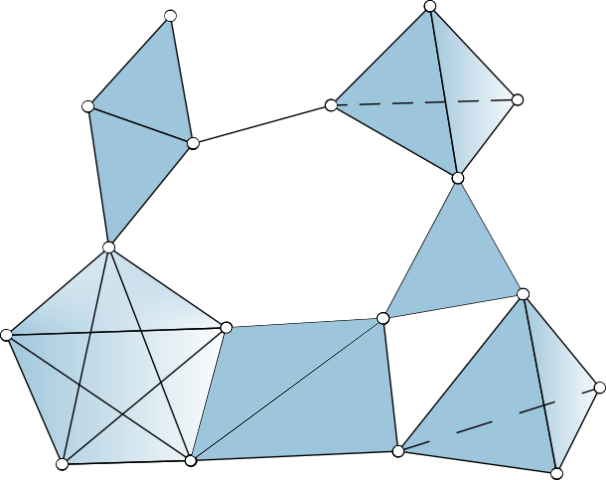}
  \caption{\textbf{2-skeleton of a 4-complex}
      in \ref{SC}. As can be seen 1-dimensional holes are preserved, while 2-dimensional holes are not, in particular there are 3 $H_2$-holes that were not in the original complex, namely those inside the two 3-simplices, and one inside the 4-simplex}
      \label{map}
      \end{figure}

As a trivial example consider the 2-skeleton of the tetrahedron, this contains a homological cycle of dimension 2, as there is a void bounded by triangles inside the tetrahedron. But no homological cycle of dimension 1 nor 0, as all its edges are in the boundary of some 2-simplex. An illustration of the 2-skeleton of a complex can be seen in Figure \ref{map}.

\section*{Experimental results}

\begin{figure}
\centering
\includegraphics[width=0.45\textwidth]{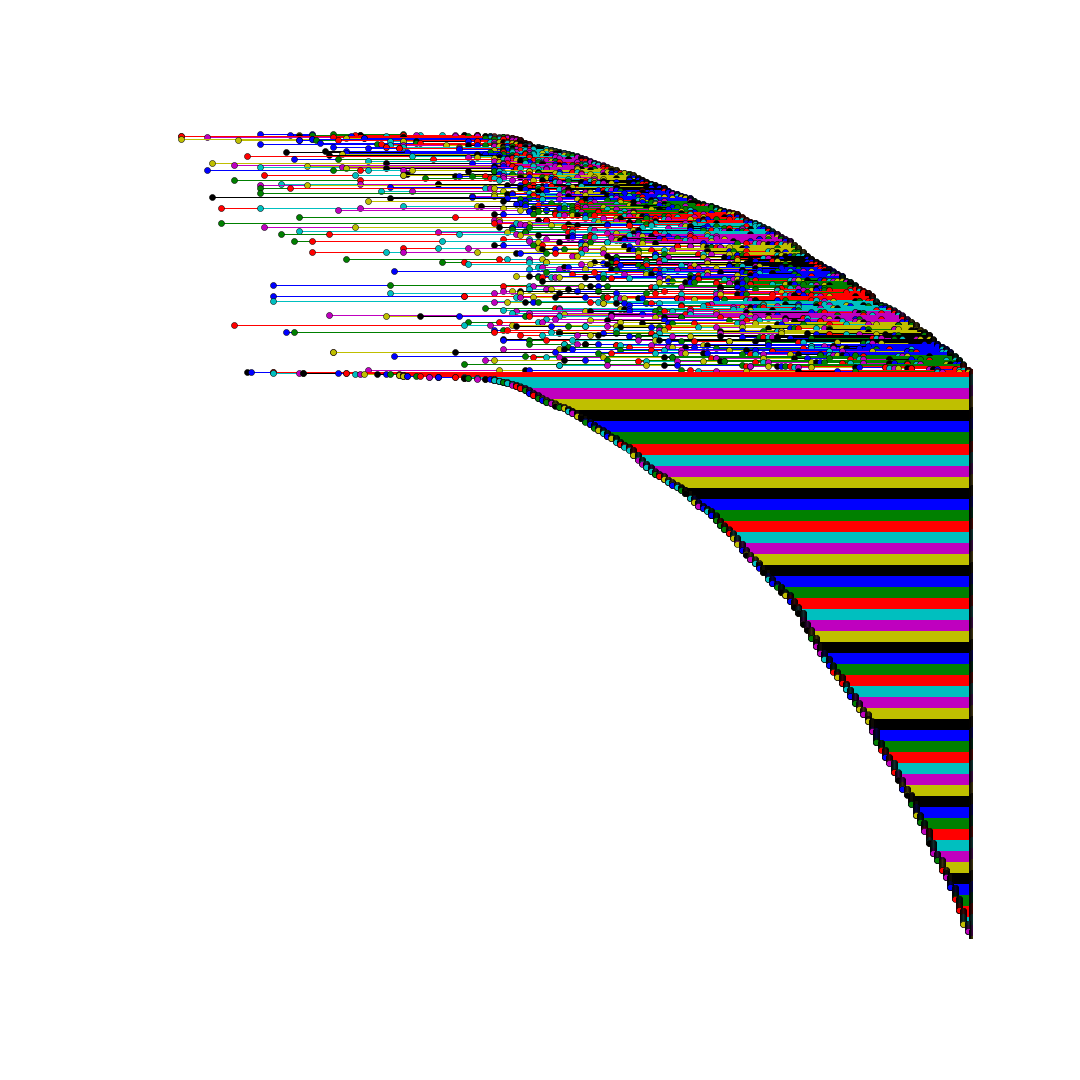}
\includegraphics[width=0.45\textwidth]{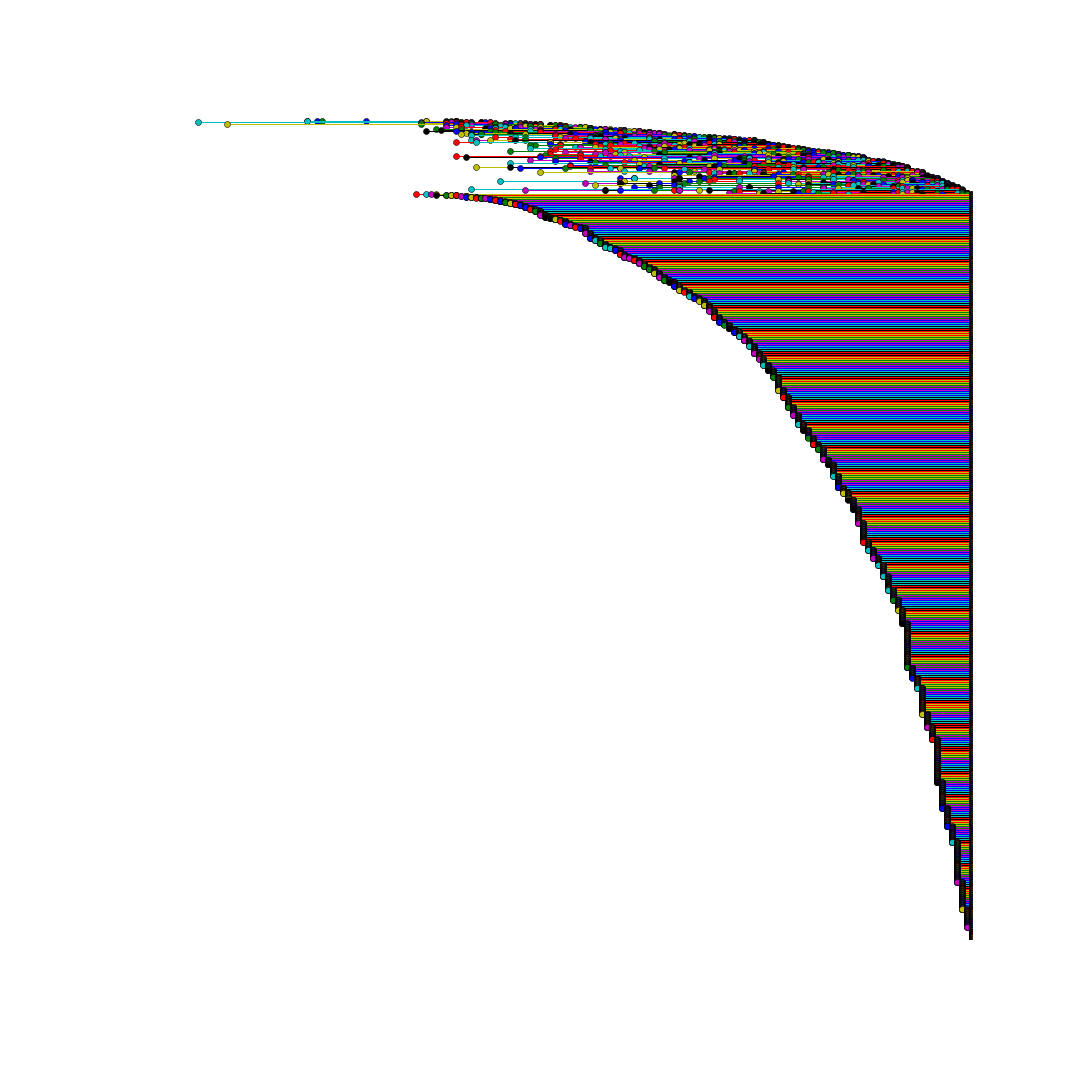}
\caption{ Barcodes of homological cycles in $H_1$ (left) and $H_2$ (right), ordered for their death time. The left and right endpoints of each segment represent the first and last step in the filtration where the hole appears.}
\label{barcodes}
\end{figure}

\begin{figure}[h!]
\centering
 \includegraphics[width=.45\textwidth]{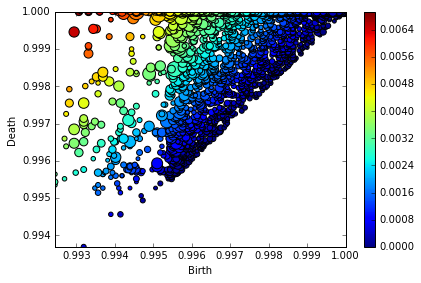}
  \includegraphics[width=.45\textwidth]{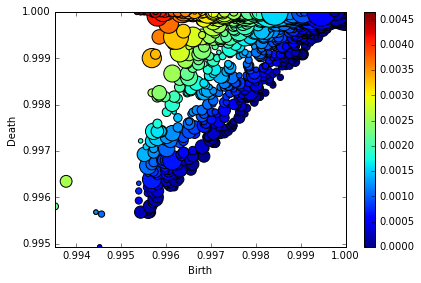}
  \caption{\textbf{Persistence diagrams for $H_1$ and $H_2$}
     Persistence diagrams report the same information of barcodes, where each barcode is mapped to a peak function that is proportional to the life of the homological feature. The heatmap reports information on the length of the cycle and the coordinates are birth and death times.}
      \label{persistence}
      \end{figure}

\begin{figure}[h!]
      \centering
 \includegraphics[width=.8\textwidth]{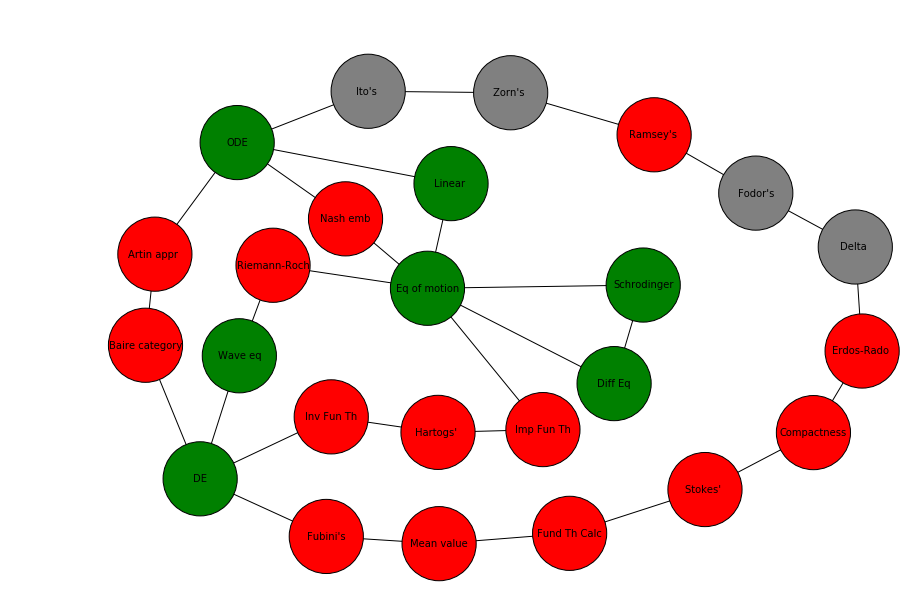} \\ 
  \includegraphics[width=0.8\textwidth]{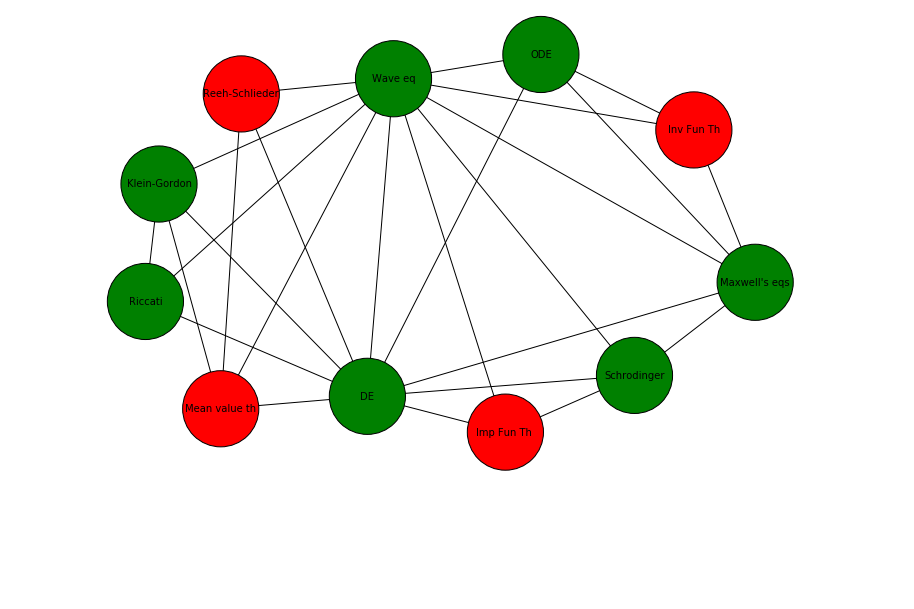}
  \caption{\textbf{ }
      Example of a $H_1$ cycle, a hole bounded by edges (top) and a $H_2$ cycle, hole bounded by triangles (bottom). Nodes color: red for theorems, grey for lemmas and green for equations. }
      \label{cycles}
      \end{figure}

We focus on homological cycles of dimension 1 and 2, respectively two dimensional holes bounded by edges and three dimensional holes bounded by triangles. Figure \ref{barcodes} reports the barcodes for $H_1$ and $H_2$, where holes are ordered by their death time (that is the step of the filtration at which they disappear), and Figure \ref{persistence} holes persistence. The first thing to notice is that most holes persist up to the end of the filtration, that is up to 03/2007, meaning that there are several areas of low connectivity (both in $H_1$ and in $H_2$) in the conceptual space, while this was not emerging from the network analysis of our data. Moreover new holes are born continuously, along all the steps of the filtration. Recall that a homological hole is a $k$-chain which is not the boundary of any higher-order structure, which means that the concepts in a $k$-cycle only appear together in sets of at most $k+1$ elements (in the $k$-simplices making the cycle). This means that a $k$-dimensional hole can be killed by an article including more than $k+1$ of the concepts in the cycle. Consider for example the $H_1$ cycle on the left of Figure \ref{cycles}, its concepts appear in the same paper in couples and at most in couples. Hence they are related but at the same time they are conceptually distant. The fact that holes continue to be born at every point in time, and most of them don't die finds a possible interpretation in that the evolution of research in mathematics proceeds by connecting new conceptual areas in a cyclic way, and rarely these concepts contribute all together to the production of scientific advances (that would kill the hole). So this suggests that the death of conceptual holes may be a sign of important advances in mathematics as the emergence of a new subfield.

\begin{figure}[h!]
             \centering
 \includegraphics[width=.8\textwidth]{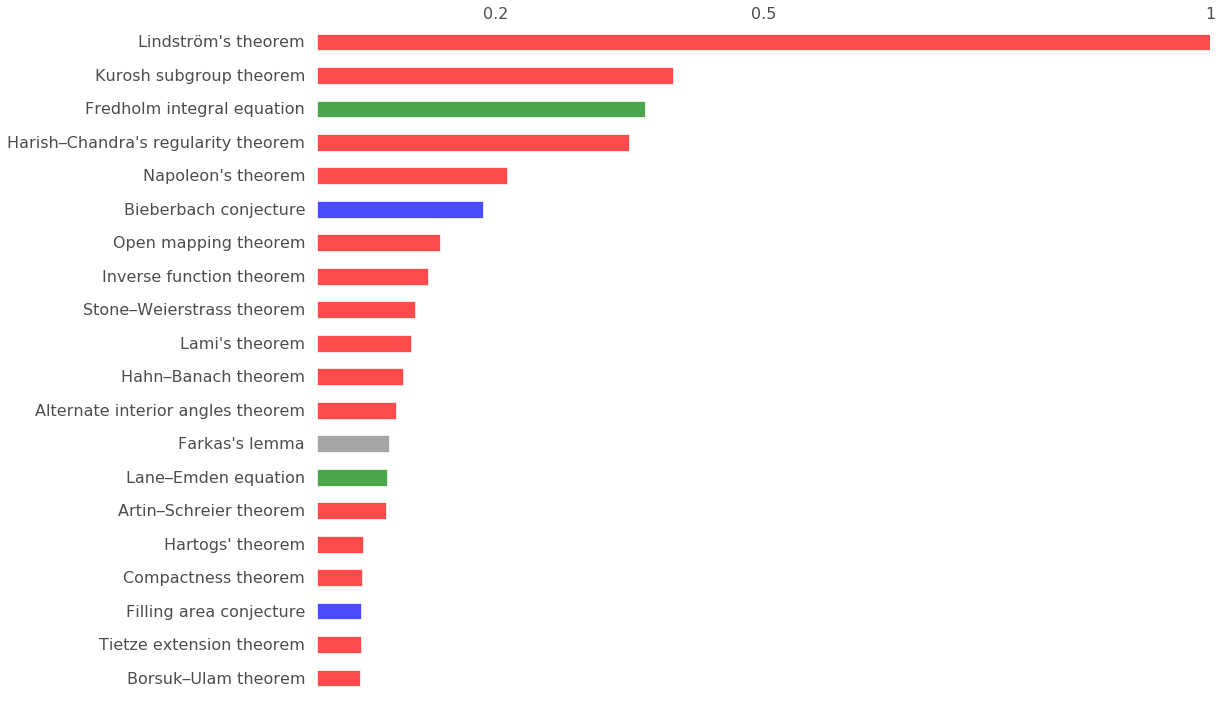} \\ 
  \includegraphics[width=.8\textwidth]{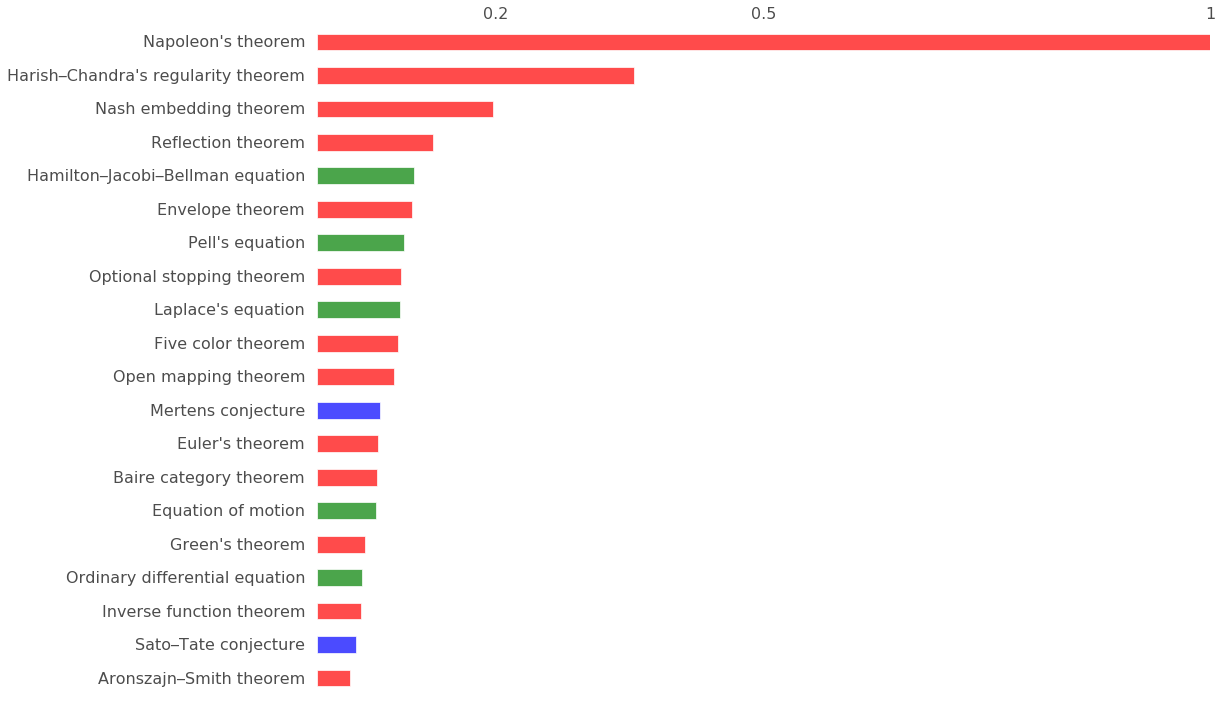}
  \caption{\textbf{Concept importance in cycles}
      20 most important concepts appearing in $H_1$ cycles (top) and $H_2$ cycles (bottom). The bars length corresponds to the number of appearances in cycles divided by the number of appearance in edges and triangles respectively.}
      \label{concH1H2}
      \end{figure}

We investigate which are the most important concepts in $H_1$ and $H_2$ by counting the number of times each concept appears in a cycle, divided by the number of times it appears in different articles, to correct for the fact that very common concepts are also more likely to appear in cycles. The results are reported in Figure \ref{concH1H2}, notice that most of the concepts are theorems there are 5 concepts in common between the most important 20 of $H_1$ and $H_2$, even if their ordering is not preserved.

\begin{figure}[h!]
   \centering
 \includegraphics[width=.45\textwidth]{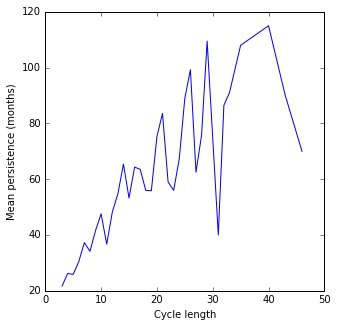}
  \includegraphics[width=.45\textwidth]{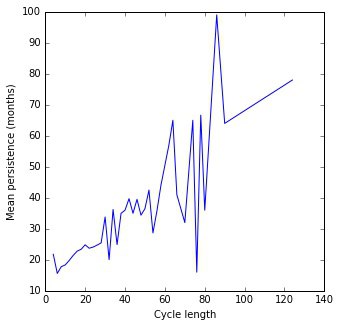}
  \caption{\textbf{Longer cycles live longer}
      The plots report cycle size on the $x$-axis and cycle mean persistence on the $y$-axis for $H_1$ (left) and $H_2$ (right). In both cases the trend shows that longer cycles are more difficult to break.}
      \label{pers_length}
      \end{figure}

A clear feature emerging from both $H_1$ and $H_2$ is that longer cycles are more difficult to break. This can be seen by plotting the length of cycles versus their average persistence, as shown in Figure \ref{pers_length}: despite some noise, a positive trend appears clearly. This suggests that concepts appearing in a long cycle that are not successive elements of the $k$-chain (so that don't appear in the same article) have a great distance among them in the conceptual space of mathematics compared to those appearing in shorter cycles. Notice that this is not trivial, as we are saying that it does not matter how many $k$-steps away in a cycle two concepts are, but only the overall length of the cycle. Moreover we would expect longer cycles to be easier to close: let a $k$-dimensional hole contain $n$ concepts, in order to kill it we need to have at least $k+1$ concepts among those $n$ to appear in an article. Assuming that each concept has the same probability of appearing in a killer article (which is of course not true), for a fixed $m > k$ the probability of having an $m$ subset from $n$ grows with growing $n$.

Exploring the death of holes more in detail, we record for each cycle which is the simplicial complex that kills it, and wherever there is more than one suspect (all those simplices appearing in the filtration when the hole dies having an intersection with the cycle at least as big as the homological dimension of the cycle) we take the simplex with largest intersection with simplices in the cycle.
Looking at the distribution of killers' dimension, the largest simplex in $H_0$ is made of 27 concepts, while both in $H_1$ and $H_2$ has  38 concepts (it is actually the same article in both). Interestingly and not surprisingly, these two articles are both surveys, the largest for $H_0$ regards open questions in number theory \cite{Waldscmidt03} and the largest for $H_1$ is a survey on differential geometry \cite{Yau06}.
The the most frequent killers' dimensions for $H_0$, $H_1$ and $H_2$ are respectively 4,7,11.

To give a better idea of the meaning of holes death, let us consider some example. The smallest cycle in $H_1$ is an empty triangle, one of such cycles is given by the three simplices \{(\emph{Schur's lemma}, \emph{Stone-Von Neumann Theorem}), (\emph{Schur's lemma}, \emph{Spectral Theorem}), (\emph{Spectral Theorem}, \emph{Stone-Von Neumann Theorem})\} which is killed by a 2-simplex when the three concepts appear together in the paper \cite{Mantoiu04}.
Another interesting example is a 5-step long cycle in $H_1$,  \{(\emph{Boltzmann equation}, \emph{Alternate Interior Angles Theorem}),
  (\emph{Boltzmann equation}, \emph{Vlasov equation}), (\emph{Inverse function theorem}, \emph{Vlasov equation}), (\emph{Arzel\'a - Ascoli theorem}, \emph{Alternate Interior Angles Theorem}), (\emph{Arzel\'a - Ascoli theorem}, \emph{Inverse function theorem})\}, that is killed by the 10-simplex which nodes are:  \{ \emph{Blum's speedup theorem}, \emph{Boltzmann equation}, \emph{Alternate Interior Angles Theorem} , \emph{Kramers theorem}, \emph{Perpendicular axis theorem}, \emph{Ordinary differential equation},\emph{ Kronecker's theorem}, \emph{Arzel\'a - Ascoli theorem}, \emph{Navier - Stokes equations}, \emph{Vlasov equation}, \emph{Moreau's theorem}\} \cite{Gottlieb00}. The article, classified in arXiv as \emph{Probability}, establishes the conditions for a family of $n$-particles Markov processes to propagate chaos, and shows its application to kinetic theory. We think this is a another possible interpretation of killing holes: a theoretical result that has several applications, hence bridges related areas and closes a homological cycle.

Simplicial stars represent potentially interesting structures in the conceptual space, which can be visualised as small substructures attached to the 'surface' of a densely connected cluster, like receptors on the membrane of a cell. To grasp the intuition consider that if we partition the concepts in a $S^k$ star between those in the core (the 0-faces of the core $k$-simplex) and those in the periphery (the zero faces of the $ (k+1)$ simplices that have the core as common face, without the core faces) by definition there can't be any edge (or higher-order simplex) between any of the concepts in the periphery. This means that periphery nodes 'touch' the surface of a densely connected area, and each of them belongs to a different simplex lying on the surface, while nodes in the core are one step far from the surface.
Considering only those with at least two simplices we count 567 $S^3$ stars and 284 $S^4$ stars. We do not check for higher-order stars for computational reasons.

\begin{figure}[h!]
         \centering
 \includegraphics[width=.8\textwidth]{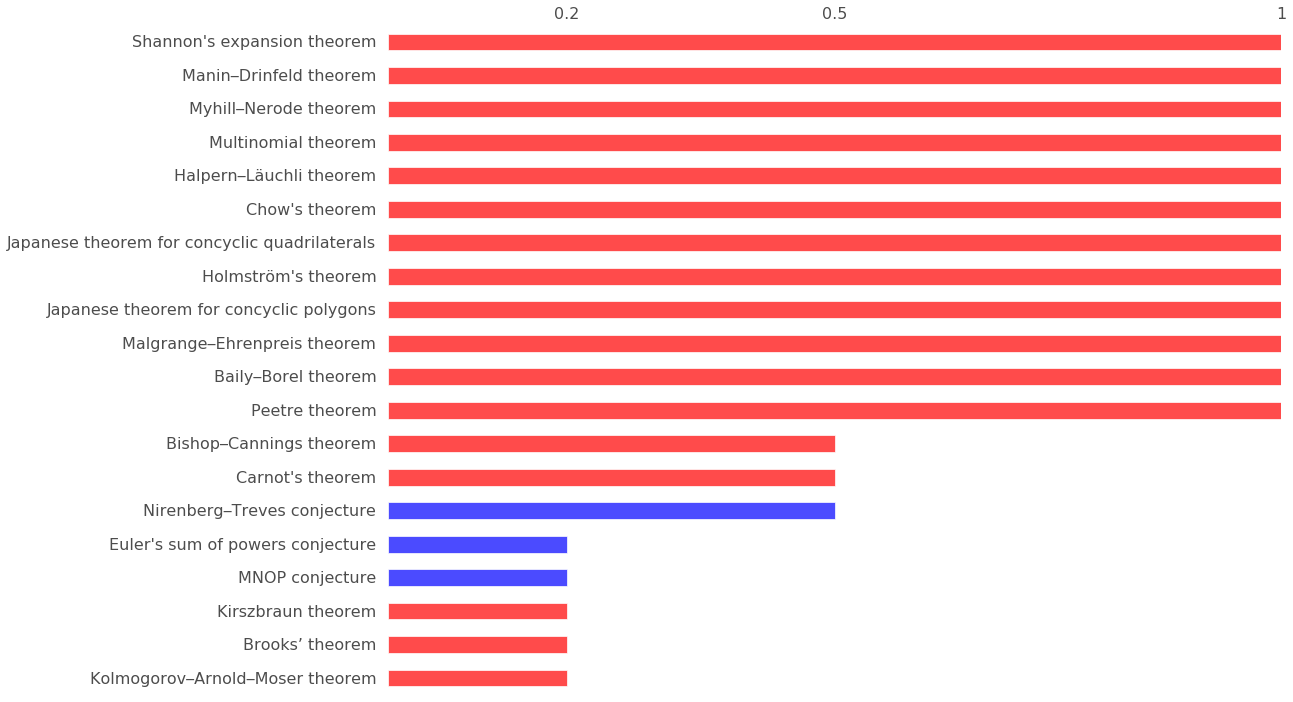}
  \includegraphics[width=.8\textwidth]{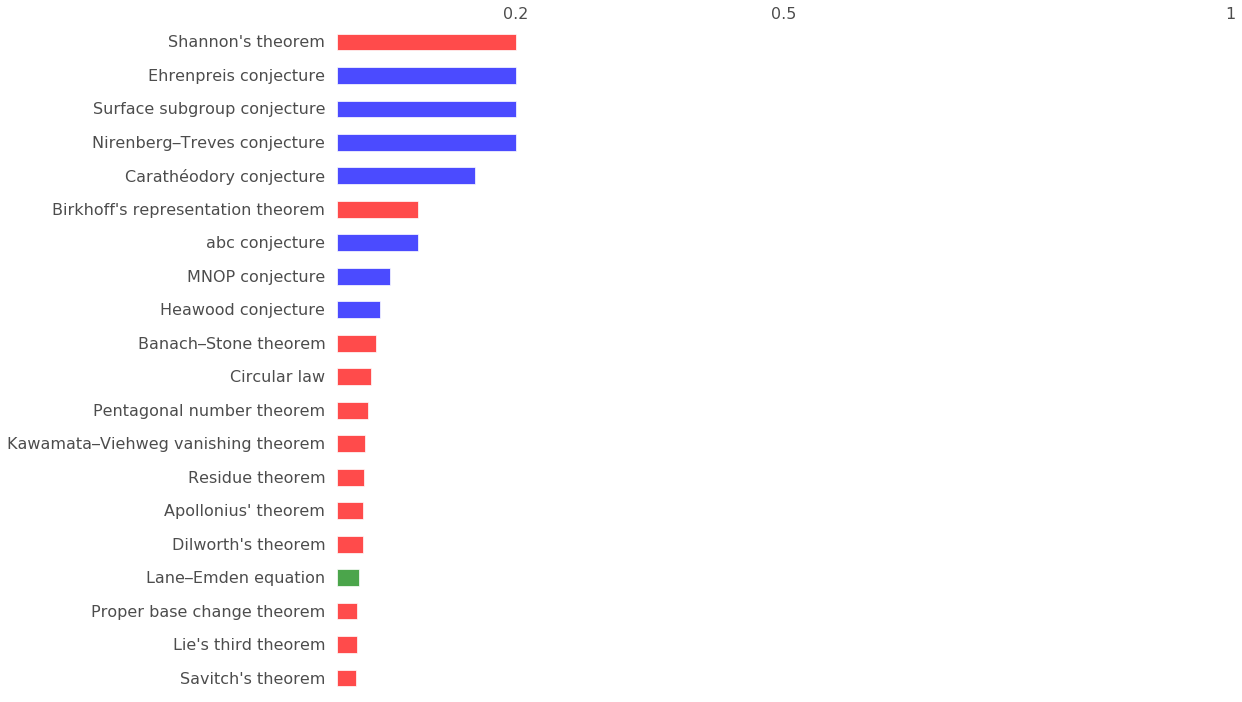}
  \caption{\textbf{Concept importance in $S^4$ stars}
      20 most important concepts appearing in the core (top) and the periphery (bottom) of $S^4$ stars. The bars reports the number of appearances in cores or peripheries divided by the number of appearance in triangles and tetrahedra respectively.}
      \label{conc4stars}
      \end{figure}

\begin{figure}[h!]
         \centering
 \includegraphics[width=.8\textwidth]{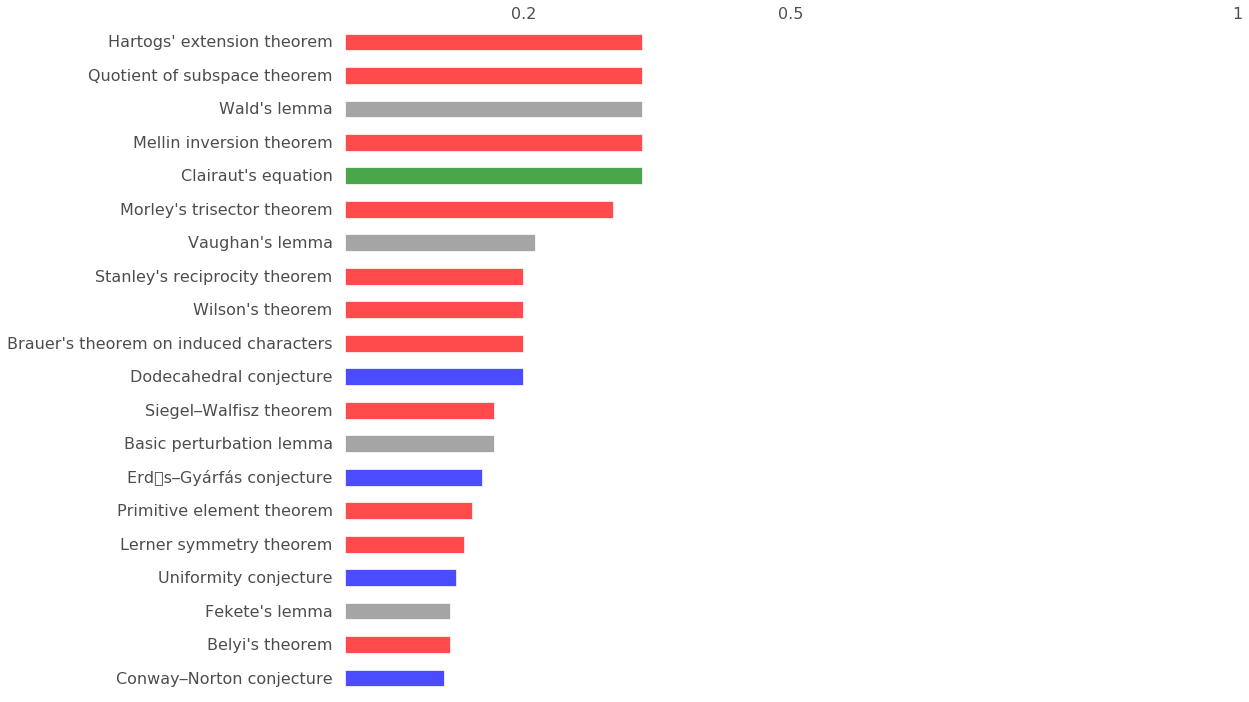}
  \includegraphics[width=.8\textwidth]{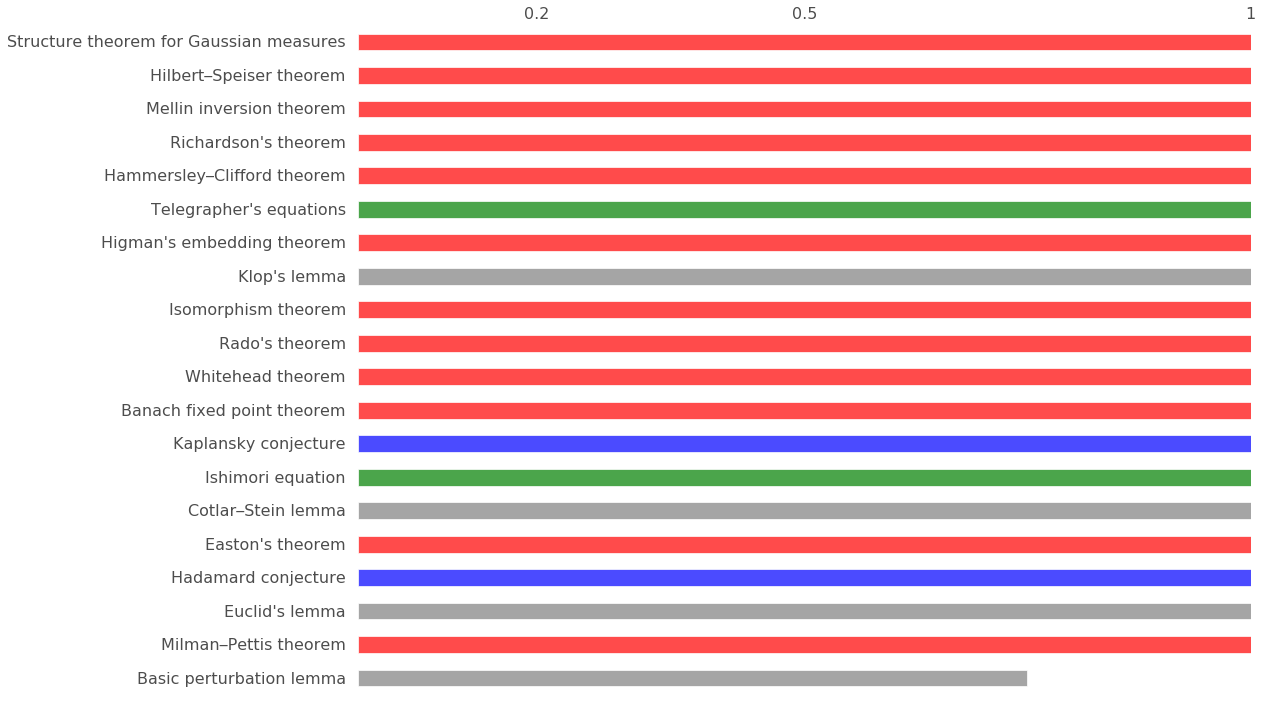}
  \caption{\textbf{Concept importance in $S^3$ stars}
      20 most important concepts appearing in the core (top) and the periphery (bottom) of $S^3$ stars. The bars reports the number of appearances in cores or peripheries divided by the number of appearance in edges and triangles respectively.}
      \label{conc3stars}
      \end{figure}

Figures \ref{conc4stars} and \ref{conc3stars} report the 20 most important concepts in the cores and peripheries of $S^4$ and $S^3$ respectively, adjusted for the number of times concepts appear in triangles (for cores) and tetrahedra (for peripheries). Notice that in both cases all except one concept are theorems/conjectures (most of them classifiable in highly abstract subfields of mathematics).  Figure \ref{concH1H2} reports the ranking of the first 20 concepts in $H_1$ and $H_2$, where here we adjust for the frequency of appearance of a concept in the simplex that constitutes the cycles, hence edges and triangles respectively. Even in these two rankings the large majority of the concepts are theorems. 

In order to check if edges that appear in cycles are also likely to appear in stars, we find the intersection between the set of concepts in stars (differentiating between edges in the core and edges in the periphery) and cycles and divide by the total numbers of edges in the corresponding cycle. As appearing in a star is a Bernoulli variable, we can easily compute the standard deviations for our estimated probabilities. Table \ref{rnd} reports the results: it is interesting that edges in the cores of both $S^3$ and $S^4$ are more likely to appear in cycles of all dimensions than edges in the peripheries. It is particularly striking that edges in the peripheries of $S^4$ never appear in any cycle. Moreover edges in stars (both cores and peripheries) are more likely to appear in cycles than a random edge, except for one case: a random edge is more likely to be in a $H_3$ cycle than an edge in the periphery of a $S^3$ star.

\begin{table}[h!]
\centering
\caption{Probabilities and standard errors for edges in the cores and peripheries of stars and of a random edge to be in $H_k$}
      \begin{tabular}{cccc}
        \hline
           & $H_1$  & $H_2$   & $H_3$\\ \hline
        $S^3$ (cores) & 0.794 (0.005) & 0.806 (0.004) & 0.785 (0.003) \\
        $S^3$ (peripheries) & 0.571 (0.006) & 0.51 (0.005) & 0.547 (0.003)\\
        $S^4$ (cores) &  0.743 (0.005) & 0.905 (0.003) & 0.91 (0.002) \\
        $S^4$ (peripheries) & 0.0 (0.0) & 0.0 (0.0) & 0.0 (0.0)\\
         Random Edge & 0.240 (0.002) & 0.300 (0.002) & 0.641 (0.003)\\
         \hline
         \label{rnd}
      \end{tabular}
\end{table}

\section*{Authors analysis}

We use the conceptual content of articles to classify the activity of researchers, by constructing for each author an activity vector:

\[
\mathbf{a}_i = (a^1_i, \dots, a^N_i)
\]
where $a^c_i $ represents the relative importance of concept $c$ in the activity of author $i$, given by the number of articles, of which $a$ is one of the authors, containing concept $c$, divided by the total number of concepts the author used in different articles, so that the sum of the entries of $\mathbf{a}_i$ is one for all authors.
As in \cite{Gurciullo15} we can use this vector to map authors' research activity on the basis of the broadness of their contribution to the concept space, as captured by the entropy:

 \[
\lambda_i = -\sum_c a^c_i \ln a^c_i
\]
$\lambda_i \ge 0 $ and is 0 only for those authors who only do research about one concept. We suggest a classification of authors based on their entropy level, we define author $i$ as \emph{specialist} if $\lambda_i<1$, \emph{polymath} if $\lambda_i>2$, and \emph{mixed} if $1 \le \lambda_i \le 2$. The choice of the thresholds is arbitrary, and it is made just for exposition's sake, hence this classification is not to be intended in a true sense of the words: a little caveat here is that this estimator of authors' activity is not very informative for those authors who published only one article. Also, an author with high entropy may be a specialist in one specific topic that has application across disciplines, hence she has a diverse range of collaborations more than being a \emph{polymath} in a strict sense. Of course we cannot disentangle which is each author's contribution to a paper, but, provided we are clear on the information conveyed by this measure of specialisation, we can make use of it to capture the relation between conceptual breadth of research (being it made by a single authors who really is a \emph{polymath} or by a research group) and homological cycles.

\begin{figure}[h!]
\centering
  \includegraphics[width=0.5\textwidth]{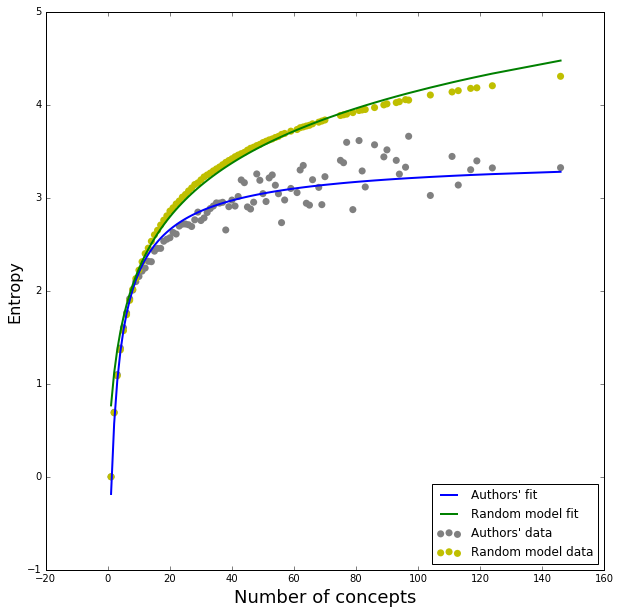}
   \caption{\textbf{Authors' entropy versus null model}
    Here we compare the entropy for authors in our dataset with average entropy over a null model. Entropy for authors is computed as the average over all authors with $k$ concepts, while entropy for the null model is the average over 1000 random extractions of $k$ concepts from our list. Average entropy for authors is always below the null model, and it is bounded above by a horizontal asymptote.}
      \label{null}
      \end{figure}

While it is expected that there is a positive relation between number of concepts in the activity vector and its entropy, we want to see how this relation compares with a null model. The null model is constructed as follows: for each concept in our list we add to an urn as many of its copies as the number of times it appears in different articles. Then for each author we count how many concepts she used across all her publications, or equivalently we sum the number of concepts used in each paper she authored (for example if an author published three papers using concept $c_1$, one using concept $c_2$ and one using concept $c_3$ her total count will be 5).  Call $k^M$ the max of these counts across all authors, then for $k \in [1, k^M]$ ($k$ integer), we extract $k$ concepts at random from the urn, and we compute the entropy of the extracted $k$-tuple, repeating the operation 1000 times for each $k$ and computing the average entropy over the 1000 extractions. To compare with authors, we group them according to the number of concepts used, and compute the average entropy for each $k$. We estimate the relation $\hat{e}$ between number of concepts $k$ and average entropy, using least squares,  finding logarithmic relation  $\hat{e}_{null} = A + B \log(1+k)$ for the null model, while for the data $\hat{e}_{data} = A + B \log(1+k)/k$ as can be seen in Figure \ref{null}. For the data $(A,B)$ is $(3.459,-5.257)$  and $(0.170, 0.863)$ for the null model, showing how, for large enough values of $k$, the fit for the data always lies below the null model, and more importantly, that the relation emerging from the data has a horizontal asymptote, while the null model does not. This is telling us that there is an upper bound on the conceptual entropy, and even very prolific \emph{polymaths} show some degree of specialisation, in the sense that as the number of articles increases, they eventually stop broadening their research including new concepts, and tend to publish new research regarding concepts they already explored.

To understand if there is any relation between authors' profile in terms of their specialisation and their contribution to homological cycles, we compute for author $i$ a measure of his \emph{homological importance} 

\[
h_i = \sum_{c \in C_i, k} \mathbf{1}_{H_k}(c)
\]
where $C_i$ is the set of concepts used by author $i$, and $\mathbf{1}_{H_k}(c)$ is the indicator function, giving 1 if concept c is in the homological cycle $H_k$.

\begin{figure}[h!]
           \centering
 \includegraphics[width=0.5\textwidth]{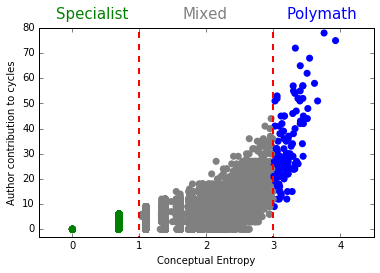}
  \caption{\textbf{Author contribution to cycles}
      The plot shows that contribution to homological cycles in $H_1$, $H_2$ and $H_3$ is positively correlated to the conceptual entropy.}
      \label{entropyc}
      \end{figure}

\begin{figure}[h!]
         \centering
  \includegraphics[width=0.5\textwidth]{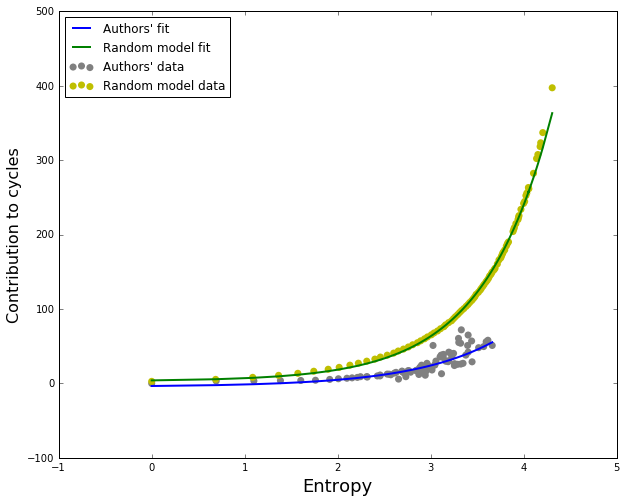}
   \caption{\textbf{Authors' contribution to cycles versus null model}
    Here we compare the contribution to homological cycles as a function of authors' entropy in our dataset with average contribution over the null model, the latter being always above the fit for authors' data.}
      \label{null1}
      \end{figure}

To correct for the fact that most frequent concepts tend to appear in cycles more often, we exclude the first 100 most frequent concepts from the computation of the homological importance. After removing the 100 most frequent concepts still the $86.5 \%$ of authors contribute at least once to a homological cycle. This confirms that homological cycles are ubiquitous, and really constitute a feature of mathematical research.
Figure \ref{entropyc} shows the scatterplot of authors' conceptual entropy and homological importance. It reveals a non-linear positive relation between the two, so more interdisciplinary authors contribute more to homological cycles, thus confirming the intuition that cycles are made by concepts that belong to different areas of mathematics, which are mostly unconnected among them. \emph{Polymaths} are often found on the boundary of these voids surrounded by concepts belonging to different conceptual areas.
In Figure \ref{null1} we show the relation $\hat{c}$ between average entropy and average importance in cycles (least squares fit), and how it compares with the null model. The relation between average entropy $\hat{\lambda}$ and average contribution to cycles is exponential both for the null model and for the data, with $\hat{c} = A + B^{\hat{\lambda}}$, and the null model always lies above the data, with $(A,B)= (3.04,3.92)$ for the null model and $(-4.50, 3.05)$ for the data.

\section*{Discussion}
In this paper, we have studied the topological structure of conceptual co-occurrences in mathematics articles, using data from arXiv. We modelled co-occurrences in a simplicial framework, focusing on higher-order relations between concepts and applying topological data analysis tools to explore the evolution of research in mathematics. We find that homological holes are ubiquitous in mathematics, appearing to show an intrinsic characteristic of how research evolves in the field: holes are likely to represent groups of concepts that are closely related but do not belong to a unitary subfield, and the death of a hole is either a sign that anticipates a potential advance in that conceptual area (for example a review trying to bridge the concepts and suggesting research lines), or an actual advance, that is an article that unify a subgroup of concepts in the cycle, for example a theoretical result with application to different areas. Less interesting, but we cannot exclude it as we have no other way to verify than reading each single paper which kills a hole, a hole-killer (especially if of very large dimension) could be a scarcely relevant article mentioning many concepts without providing any true contribution.  

We also find that the higher the number of concepts in a hole, the longer it takes to die, hence the length of a hole is a good proxy of how distant these concepts are, in terms of their likelihood to appear together in an article. So in this sense large holes could be seen as potential spaces for important advances in mathematics. Moreover we further explore the structure of co-occurrences by looking at the simplicial analogs of stars in higher dimension, which represent groups of concept (those in the core of the star) that supports and connects many otherwise unrelated concepts, and we find that concepts appearing in stars tend also to appear in holes more often than they would do at random, suggesting that both structures lie at the frontier of mathematical research.

We also explore authors' conceptual profile by ordering them on the basis of their conceptual entropy, so that we can differentiate between those authors who tend to specialise and publish mostly about few concepts, and others that do research on a broad range of topics, that we call \emph{polymaths}. Comparing authors' profiles with a random model, we find that authors' entropy as a function of how may concepts authors use across different publications, is bounded above, while in the null model, entropy is always increasing for larger set of random concepts. This is reasonable, and means that even the more prolific \emph{polymaths}, even if they publish a large number of articles, will still tend to specialise to some extent, instead of doing research always on new topics. Moreover we find that \emph{polymaths} contribute to homological holes more than specialists, so \emph{polymaths} are often at the frontier of research.

This work is a first attempt to explore the importance of homological holes in mathematics, and it is important to issue certain caveats here. It seems clear that our observations would not have been possible within a classical network analysis of co-occurrences, and we also think that holes deaths can  be informative in capturing advances in the discipline.
However, we do not claim that we are describing the essence of mathematical practice here. If on one hand, by extracting concepts from the whole text of the article instead of just focusing on the keywords we avoid the bias of authors' own classification of their work, on the other hand it is possible that for some articles the conceptual content is not well captured by our approach: we cannot exclude that the set of concepts that better identify the content of the article do not find an exact match in our list, while those finding a match are poorly representative of the article. We believe that such cases are a minority as our list is very comprehensive, still this is an aspect to take in consideration when analysing our results, even if this is an issue that is potentially arising every time one does text analysis, irrespective of the tools used to explore the data. In this direction, an important step would be to assess the robustness of our observations by purposely introducing errors in our data analysis, for instance by focusing only on a fraction of our list of identified concepts, and keeping the other unknown.

Overall, this paper suggests new directions for the study of co-occurrences by focusing on their higher-order properties and there is substantial space for further development. The first and most urgent point, in our opinion, regards the necessity to validate the findings from the data by having well-founded null models for comparison. Recently some null models for simplicial complexes have been developed \cite{Young17, Courtney16}, which we did not use here for computational reasons, as producing samples from our relatively large dataset proved too challenging to manage with our resources. Moreover it would be interesting to develop a simplicial complex formation model able to reproduce the features of the evolution of research as it emerges from data. Further work could be done by using larger datasets, as it would be very interesting to explore the birth and death of holes in a larger time-span, and to study simplicial co-occurrences in other disciplines, in order to see if any difference appears in the way research evolves in different fields. Furthermore, conceptual spaces emerging from co-occurrences relations could be explored adding a further dimension to the filtration: in our case we focus on a temporal filtration, disregarding the weights of simplices, this could be extended by filtering along time and weight using multidimensional persistence \cite{CarlssonZomo09}.

\section*{Acknowledgement}
The authors thank Oliver Vipond and Alexey Medvedev for useful discussions and suggestions. DC received support from FNRS (Belgium).

\bibliographystyle{acm}
\bibliography{metamath_biblio}

\begin{thebibliography}{10}

\bibitem{Akimushkin17}
{\sc Akimushkin, C., Amancio, D.~R., and Oliveira, O. N.~J.}
\newblock Text authorship identified using the dynamics of word co-occurrence
  networks.
\newblock {\em PLos ONE 12}, 1 (2017).
\newblock 1-101.

\bibitem{Carlsson09}
{\sc Carlsson, G.}
\newblock Topology and data.
\newblock {\em Bullettin AMS 46}, 2 (2009), 255--308.

\bibitem{CarlssonZomo09}
{\sc Carlsson, G., and Zomorodian, A.}
\newblock The theory of multidimensional persistence.
\newblock {\em Discrete \& Computational Geometry 42\/} (2009), 71--93.

\bibitem{Carlsson05}
{\sc Carlsson, G., Zomorodian, A., Collins, A., and Guibas, L.~J.}
\newblock Persistence barcodes for shapes.
\newblock {\em International Journal of Shape Modeling 11\/} (2005), 149--188.

\bibitem{Carstens13}
{\sc Carstens, C.~J., and Horadam, K.~J.}
\newblock Persistent homology of collaboration networks.
\newblock {\em Mathematical Problems in Engineering\/} (2013).
\newblock http://dx.doi.org/10.1155/2013/815035.

\bibitem{Chan13}
{\sc Chan, J.~M., Carlsson, G., and Rabadan, R.}
\newblock Topology of viral evolution.
\newblock {\em PNAS 110}, 46 (2013).

\bibitem{Chiang07}
{\sc Chiang, I.~J.}
\newblock Discover the semantic topology in high-dimensional data.
\newblock {\em Expert Systems with Applications 33\/} (2007), 256--262.

\bibitem{Courtney16}
{\sc Courtney, O., and Bianconi, G.}
\newblock Generalized network structures: The configuration model and the
  canonical ensemble of simplicial complexes.
\newblock {\em Physical Review E 93\/} (2016).

\bibitem{Edelsbrunner08}
{\sc Edelsbrunner, H., and Harer, J.}
\newblock Persistent homology - a survey.
\newblock {\em Contemporary Mathematics 453}, 2 (2008), 255--308.

\bibitem{Estrada17}
{\sc Estrada, E., and Ross, G.}
\newblock Centralities in simplicial complexes.
\newblock {\em arXiv preprint arXiv:1703.03641\/} (2017).

\bibitem{Ferrer01}
{\sc Ferrer-i Cancho, R., and Sol\'e, R.~V.}
\newblock The small world of human language.
\newblock {\em Proc. R. Soc. Lond. B 268\/} (2001), 2261--2265.
\newblock DOI: 10.1098/rspb.2001.1800.

\bibitem{Garg18}
{\sc Garg, M., and Kumar, M.}
\newblock Identifying influential segments from word co-occurrence networks
  using ahp.
\newblock {\em Cognitive Systems Research 47\/} (2018), 23--41.

\bibitem{Gottlieb00}
{\sc Gottlieb, A.~D.}
\newblock Markov transitions and the propagation of chaos.
\newblock {\em arXiv:math/0001076\/} (2004).

\bibitem{Gurciullo15}
{\sc Gurciullo, S., et~al.}
\newblock Complex politics: A quantitative semantic and topological analysis of
  uk house of commons debates.
\newblock {\em arXiv preprint arXiv:1510.03797\/} (2015).

\bibitem{Horak09}
{\sc Horak, D., Maletic, S., and Rajkovic, M.}
\newblock Persistent homology of complex networks.
\newblock {\em Journal of Statistical Mechanics: Theory and Experiment 3\/}
  (2009), 3--34.

\bibitem{Jenssen01}
{\sc Jenssen, T.-K., Laegreid, A., Komorowski, J., and Hovig, E.}
\newblock A literature network of human genes for high-throughput analysis of
  gene expression.
\newblock {\em Nature Genetics 28\/} (2001), 21--28.

\bibitem{Jo11}
{\sc Jo, Y., Hopcroft, J.~E., and Lagoze, C.}
\newblock The web of topics: Discovering the topology of topic evolution in a
  corpus.
\newblock In {\em Proceedings of the 20th international conference on World
  wide web, March 28 - April 01, 2011, Hyberabad, India\/} (2011), ACM, New
  York, NY, USA, pp.~256--266.

\bibitem{Lazer09}
{\sc Lazer, D., Mergel, I., and Friedman, A.}
\newblock Co-citation of prominent social network articles in sociology
  journals: The evolving canon.
\newblock {\em Connections 29}, 1 (2009).

\bibitem{Mamuye16}
{\sc Mamuye, A.~L., Rucco, M., Tesei, L., and Merelli, E.}
\newblock Persistent homology analysis of rna.
\newblock {\em Mol. Based Math. Biol 4\/} (2016), 14--25.

\bibitem{Mantoiu04}
{\sc Mantoiu, M., Purice, R., and Richard, S.}
\newblock Twisted crossed products and magnetic pseudodifferential operators.
\newblock {\em arXiv:math-ph/0403016\/} (2004).

\bibitem{Mullen14}
{\sc Mullen, E.~K., et~al.}
\newblock Gene co-citation networks associated with worker sterility in honey
  bees.
\newblock {\em BMC Systems Biology 8}, 38 (2014).
\newblock https://doi.org/10.1186/1752-0509-8-38.

\bibitem{Otter17}
{\sc Otter, N., Porter, M.~A., Tillmann, U., Grindrod, P., and Harrington, H.}
\newblock A roadmap for the computation of persistent homology.
\newblock {\em EPJ Data Science 6}, 17 (2017).
\newblock https://doi.org/10.1140/epjds/s13688-017-0109-5.

\bibitem{Pal17}
{\sc Pal, S., Moore, T.~J., Ramanathan, R., and Swami, A.}
\newblock Comparative topological signatures of growing collaboration networks.
\newblock In {\em Proceedings of the 8th Conference on Complex Networks
  CompleNet 2017\/} (2017), R.~S. B~Gon\c{c}alves, R~Menezes and V.~Zlatic,
  Eds., Stoneham: Butterworth-Heinemann, pp.~16--27.

\bibitem{Patania17}
{\sc Patania, A., Petri, G., and Vaccarino, F.}
\newblock The shape of collaborations.
\newblock {\em EPJ Data Science 6}, 18 (2017).
\newblock DOI 10.1140/epjds/s13688-017-0114-8.

\bibitem{Patania17_2}
{\sc Patania, A., Petri, G., and Vaccarino, F.}
\newblock Topological analysis of data.
\newblock {\em EPJ Data Science 6}, 7 (2017).
\newblock DOI 10.1140/epjds/s13688-017-0104-x.

\bibitem{Petri14}
{\sc Petri, G., Expert, P., Turkheimer, F., Carhart-Harris, R., Nutt, D.,
  Hellyer, P.~J., and Vaccarino, F.}
\newblock Homological scaffolds of brain functional networks.
\newblock {\em Journal of the Royal Society Interface 11}, 20140873 (2014).
\newblock http://dx.doi.org/10.1098/rsif.2014.0873.

\bibitem{Petri13}
{\sc Petri, G., Scolamiero, M., I, D., and F, V.}
\newblock Topological strata of weighted complex networks.
\newblock {\em PLoS ONE 8}, 6 (2013).
\newblock doi:10.1371/ journal.pone.0066506.

\bibitem{Radhakrishnan17}
{\sc Radhakrishnan, S., Erbis, S., Isaacs, J.~A., and Kamarthi, S.}
\newblock Novel keyword co-occurrence network-based methods to foster
  systematic reviews of scientific literature.
\newblock {\em PLos ONE 12}, 3 (2017), 23--41.
\newblock e0172778. https://doi.org/10.1371/journal. pone.0172778.

\bibitem{Sami17}
{\sc Sami, I.~R., and Farrahi, K.}
\newblock A simplified topological representation of th text for local and
  global context.
\newblock In {\em Proceedings of the 2017 ACM on Multimedia Conference,
  Mountain View, California, USA\/} (2017), ACM, New York, NY, USA,
  pp.~1451--1456.

\bibitem{Serrano09}
{\sc Serrano, M.~A., Bogu\~na, M., and Vespignani, A.}
\newblock Extracting the multiscale backbone of complex weighted networks.
\newblock {\em PNAS 106\/} (2009), 6483--6488.

\bibitem{Slater09}
{\sc Slater, P.~B.}
\newblock A two-stage algorithm for extracting the multiscale backbone of
  complex weighted networks.
\newblock {\em PNAS 106}, 26 (2009).
\newblock https://doi.org/10.1073/pnas.0904725106.

\bibitem{Stolz17}
{\sc Stolz, B.~J., Harrington, H.~A., and Porter, M.~A.}
\newblock Persistent homology of time-dependent functional networks constructed
  from coupled time series.
\newblock {\em Chaos 27}, 047410 (2017).
\newblock https://doi.org/10.1063/1.4978997.

\bibitem{Su10}
{\sc Su, H.-N., and Lee, P.-C.}
\newblock Mapping knowledge structure by keyword co-occurrence: a first look at
  journal papers in technology foresight.
\newblock {\em Scientometrics 85\/} (2010), 65--79.
\newblock DOI 10.1007/s11192-010-0259-8.

\bibitem{Taylor15}
{\sc Taylor, D., et~al.}
\newblock Topological data analysis of contagion maps for examining spreading
  processes on networks.
\newblock {\em Nature Communications 6}, 7723 (2015).
\newblock doi:10.1038/ncomms8723.

\bibitem{Wagner12}
{\sc Wagner, H., et~al.}
\newblock Computational topology in text mining.
\newblock In {\em Computational Topology in Image Context\/} (2012), M.~Ferri,
  P.~Frosini, C.~Landi, A.~Cerri, and B.~Di~Fabio, Eds., Springer Berlin
  Heidelberg, pp.~68--78.

\bibitem{Waldscmidt03}
{\sc Waldschmidt, M.}
\newblock Open diophantine problems.
\newblock {\em arXiv:math/0312440\/} (2003).

\bibitem{Wang11}
{\sc Wang, X., Zhang, X., and Xu, S.}
\newblock Patent co-citation networks of fortune 500 companies.
\newblock {\em Scientometrics 88}, 3 (2011), 761--770.

\bibitem{Yau06}
{\sc Yau, S.-T.}
\newblock Perspectives on geometric analysis.
\newblock {\em arXiv:math/0602363\/} (2006).

\bibitem{Young17}
{\sc Young, J.-C., Petri, G., Vaccarino, F., and Patania, A.}
\newblock Construction of and efficient sampling from the simplicial
  configuration model.
\newblock {\em Physical Review E 96\/} (2017).

\bibitem{Zhang12}
{\sc Zhang, J., Xie, J., Hou, W., Tu, X., Xu, J., et~al.}
\newblock Mapping the knowledge structure of research on patient ahderence:
  knowledge domain visualization based co-word analysis and social network
  analysis.
\newblock {\em PLoS ONE 7}, 4 (2012), e34497.
\newblock doi:10.1371/journal.pone.0034497.

\end{thebibliography}

\end{document}